\documentstyle[epsf,psfig]{mn}

\newif\ifAMStwofonts
\ifoldfss
  \ifCUPmtlplainloaded \else
    \NewTextAlphabet{textbfit} {cmbxti10} {}
    \NewTextAlphabet{textbfss} {cmssbx10} {}
    \NewMathAlphabet{mathbfit} {cmbxti10} {} % for math mode
    \NewMathAlphabet{mathbfss} {cmssbx10} {} %  "   "    "
  \fi
  \ifAMStwofonts
    \ifCUPmtlplainloaded \else
      \NewSymbolFont{upmath} {eurm10}
      \NewSymbolFont{AMSa} {msam10}
      \NewMathSymbol{\upi}     {0}{upmath}{19}
      \NewMathSymbol{\umu}     {0}{upmath}{16}
      \NewMathSymbol{\upartial}{0}{upmath}{40}
      \NewMathSymbol{\leqslant}{3}{AMSa}{36}
      \NewMathSymbol{\geqslant}{3}{AMSa}{3E}

       \let\ge=\geqslant
    \fi
  \fi
\fi % End of OFSS

\ifnfssone
  \newmathalphabet{\mathit}
  \addtoversion{normal}{\mathit}{cmr}{m}{it}
  \addtoversion{bold}{\mathit}{cmr}{bx}{it}
  \newmathalphabet{\mathbfit} % math mode version of \textbfit{..}
  \addtoversion{normal}{\mathbfit}{cmr}{bx}{it}
  \addtoversion{bold}{\mathbfit}{cmr}{bx}{it}
  \newmathalphabet{\mathbfss} % math mode version of \textbfss{..}
  \addtoversion{normal}{\mathbfss}{cmss}{bx}{n}
  \addtoversion{bold}{\mathbfss}{cmss}{bx}{n}
  \ifAMStwofonts
    \ifCUPmtlplainloaded \else
      \UseAMStwoboldmath
      \makeatletter
      \new@mathgroup\upmath@group
      \define@mathgroup\mv@normal\upmath@group{eur}{m}{n}
      \define@mathgroup\mv@bold\upmath@group{eur}{b}{n}
      \edef\UPM{\hexnumber\upmath@group}
      \new@mathgroup\amsa@group
      \define@mathgroup\mv@normal\amsa@group{msa}{m}{n}
      \define@mathgroup\mv@bold\amsa@group{msa}{m}{n}
      \edef\AMSa{\hexnumber\amsa@group}
      \makeatother
      \mathchardef\upi="0\UPM19
      \mathchardef\umu="0\UPM16
      \mathchardef\upartial="0\UPM40
      \mathchardef\leqslant="3\AMSa36
      \mathchardef\geqslant="3\AMSa3E

       \let\ge=\geqslant
    \fi
  \fi
\fi % End of NFSS release 1

\ifnfsstwo
  \DeclareMathAlphabet{\mathbfit}{OT1}{cmr}{bx}{it}
  \SetMathAlphabet\mathbfit{bold}{OT1}{cmr}{bx}{it}
  \DeclareMathAlphabet{\mathbfss}{OT1}{cmss}{bx}{n}
  \SetMathAlphabet\mathbfss{bold}{OT1}{cmss}{bx}{n}
  \ifAMStwofonts
    \ifCUPmtlplainloaded \else
      \DeclareSymbolFont{UPM}{U}{eur}{m}{n}
      \SetSymbolFont{UPM}{bold}{U}{eur}{b}{n}
      \DeclareSymbolFont{AMSa}{U}{msa}{m}{n}
      \DeclareMathSymbol{\upi}{0}{UPM}{"19}
      \DeclareMathSymbol{\umu}{0}{UPM}{"16}
      \DeclareMathSymbol{\upartial}{0}{UPM}{"40}
      \DeclareMathSymbol{\leqslant}{3}{AMSa}{"36}
      \DeclareMathSymbol{\geqslant}{3}{AMSa}{"3E}

       \let\ge=\geqslant
    \fi
  \fi
\fi % End of NFSS release 2

\ifCUPmtlplainloaded \else
  \ifAMStwofonts \else % If no AMS fonts
    \def\upi{\pi}
    \def\umu{\mu}
    \def\upartial{\partial}
  \fi
\fi

\title{Hard X-Ray Lightcurves of High Mass X-Ray Binaries}
\author[S. Laycock et al.]
       {S.~Laycock,$^1$ M. J.~Coe,$^1$
       \newauthor
       C.A. Wilson$^2$, B. A.~Harmon$^2$, M.~Finger,$^2$ \\
       $^1$Department of Physics and Astronomy, Southampton University, SO17 
1BJ, UK\\
       $^2$NASA, MSFC, Huntsville, AL35812, USA}
\date{Accepted .
      Received ;
      in original form
      }

\pagerange{\pageref{firstpage}--\pageref{lastpage}}
\pubyear{2002}

\begin{document}

\maketitle

\label{firstpage}

\begin{abstract}

Using the 9 years of continuous data now available from the Burst And
Transient Source Experiment (BATSE) aboard CGRO, we have measured
orbital periods and produced folded lightcurves for 8 High Mass X-ray
Binaries (HMXB).  Given the length of the datasets, our determinations
are based on many more binary orbits than previous
investigations. Thus our source detections have high statistical
significance and we are able to follow long-term trends in X-ray
output. In particular we focus on two systems: A0538-668 and
EXO2030+375 both HMXBs exhibiting Type I outbursts.

Recent work on A0538-668 \cite{al} reported a 16.65d optical
variability due to the orbital period, but only seen during minima of
a longer-term variability at 421d. We searched for this signal in the
BATSE dataset using an ephemeris derived from Alcock et al
\shortcite{al} \& Skinner \shortcite{sk}. We found no evidence for
such modulation and place an upper limit of 3.0 $\times 10^{-3}$
photon cm$^{-2}$ s$^{-1}$ in the 20-70 keV BATSE energy band , based
upon statistical modelling of the signal.  Previous observations of
EXO 2030+375 \cite{rei} using RXTE ASM data indicate secondary
outbursts occur at apastron passage. We present a lightcurve for an
earlier epoch showing convincing evidence for such apastron
outbursts. We find apastron outbursts in 3 sources, all having orbital
periods greater than 41d. No such signal is conclusively detected in
the more rapidly orbiting systems studied.

\end{abstract}

\begin{keywords}
stars - X-rays: binaries :pulsars
\end{keywords}

\section{Introduction}

The Burst And Transient Source Experiment on the Compton Gamma Ray
Observatory \cite{fish} provided near continuous monitoring of the
whole sky at hard X-ray energies.  BATSE only sees the high energy
tail of HMXBs, typically detected in the lowest DISCLA channel and
CONT channels 1-4 covering 20-70keV.  In studying X-ray pulsars two
modes of operation are possible; pulsed flux and Earth occultation. If
pulsations are detected corresponding to the spin period of a known
pulsar then the pulsed flux and pulse frequency can be measured with
Fourier analysis or epoch folding (Bildsten, L. et al. 1997) and used
to monitor the source with great sensitivity and fine time
resolution. \cite{bil} For the long term study of behavior related to
orbital motion in binary systems, daily averages of the total flux are
sufficient and are in addition sensitive to non-pulsed X-ray output
from the source. By exploiting the Earth's limb as an occulting mask,
individual X-ray sources can be isolated with an accuracy of
~1$\degr$. Termed the Earth Occultation Technique, two measurements of
the total flux from a source are made during each 90 minute satellite
orbit as the source moves into and out of eclipse due to the
satellite's motion, full details are given by Zhang et al
\shortcite{zha} \& Harmon et al \shortcite{har}.

	In this paper we present folded X-ray lightcurves of Type I
(normal) outbursts from 8 HMXBs in which periodic fluctuations
corresponding to previously published orbital periods were
significantly detected.  Such outbursts are thought to occur at each
periastron passage of the neutron star. These outbursts last for 8-10
days and the X-ray luminosity increases by a factor of $\sim$10.  In
contrast, Type II are giant outbursts which do not correlate with the
orbital phase but tend to last for several orbital cycles. The X-ray
luminosity during the outburst may reach close to the Eddington
luminosity.

In addition we studied the 30.5d modulation
from LMC X-4 which was also a significant detection and investigated
the X-ray behaviour of A0538-668 in the light of recent optical work
\cite{al}. The BATSE datasets for these systems cover many
orbital cycles and hence present the first ever opportunity to study
their long term behavior. Previously only widely spaced, scheduled
or chance obsevations were available for this purpose. Usually
made with different instruments having little or no overlap in
terms of spectral response, extrapolating long term trends from
previous data was of limited value.

	The publicly available pulsed flux monitor reports produced
by MSFC demonstrates that HMXBs typically exhibit active epochs
lasting several months interspersed with quiescent periods. A
question we addressed was that of possible changes in behaviour
between these epochs. If a source is bright enough or its active
epochs span enough orbital cycles then folded X-ray lightcurves
from different epochs can be directly compared and variations in
behavior detected.

	Variability in the X-ray recurrence period can be looked for in
the power spectrum of the flux history. Width of the recurrence
frequency peak will be increased if that frequency varies, in other
words if the exact time at which the outbursts turn on varies from
orbit to orbit then this will show up in the power spectrum. If
such jitter can be identified this could provide clues about the
underlying physical behaviour of such systems. For example theory
indicates that accretion rates are effected by a neutron star's
speed relative to its surroundings. The Alfven speed in stellar
wind plasma is a function of both magnetic field strength and
plasma density. Thus a high outflow  velocity of a wind or
circumstellar disk may inhibit accretion during periastron passage
and push the X-ray outburst to a later phase. Additionally the
orientation of the magnetic field frozen into a plasma strongly
affects its coupling to a magnetosphere moving through it. These
effects, although extremely difficult to measure directly, may
leave a signature in possible chaotic fluctuations or jitter in
the exact phase at which outbursts turn on. The jitter is expected
to be chaotic because there is no direct connection between the
plasma properties and the binary motion, yet the stability of the
accretion disk depends on a dynamic balance between accretion rate
onto the disk and accretion rate onto the neutron star.

	Secondary outbursts occuring close to apastron of the
neutron star are also a key observation in this study. Their
existence and exact phase at which they occur will yield
information about circumstellar disk outflow and stellar wind
parameters, principally outflow velocity and density information.
Higher outflow velocity is expected to delay periastron outburst
turn-on and is cited as a possible cause of apastron outburst in
the Bondi-Hoyle theory of accretion. Our folded lightcurves show that
EXO2030+375, GX301-2 and 4U1145-619 may exhibit X-ray emision at
apastron passage of the neutron star. Whether this emission
occurs always or only sporadically, is impossible to determine from
individual orbits, so long-term coverage is a necessity.

	In these preliminary interpretations of our lightcurves
we are aware that BATSE only sees the high energy tails of the
emitted X-ray spectra of these sources. Lightcurve shape is thus
affected by the true spectral profile of the emitted X-rays and
the spectral response of the detector. The most significant effect
is expected to occur at the peak of periastron outbursts, when
there is the highest column density of obscuring matter and the
most intense flux of hard photons, leading to X-ray reprocessing
and re-radiation at lower energies.

	Parameters for the 8 systems studied are shown in Table 1.
Of these sources, in all cases but LMC X-4 the modulation studied
is directly linked to orbital motion. In LMC X-4 the binary period
of 1.4d is not seen because it is too faint in hard X-rays.
The 1.4 day orbital period of LMC X-4 is above the Nyquist limit for
this data (about 2.4d) so aliasing will shift this frequency
in the power spectrum, but no significant signal is detected.
However the well known periodicity near 30.5 days was detected and
a lightcurve constructed. An improved measurement was made of X-ray
modulation due to the precession of a warped accretion disk.

Note that throughout this paper dates are presented as Truncated Julian
Dates (TJD), where TJD = JD - 2440000.5.

\section[]{Data processing and data artefacts}

The BATSE dataset used in this work consists of daily averages
of the flux levels derived from the occultation technique at MSFC
\cite{har}.

In the occultation technique, source fluxes are in principle measured
twice every spacecraft orbit, as rising and setting steps. However, in
practice, not all occultation steps are available for measurement.
The most common reason steps are unavailable are (1) passage of the
spacecraft through the lower Van Allen radiation belt at the South
Atlantic Anomaly, when the detector high voltage was turned off; (2)
periods of time when CGRO was out of line-of-sight contact with the
NASA Tracking and Relay Satellites (TDRS); and (3) data which were
flagged by the BATSE operations team (i.e. during gamma ray bursts,
solar flares, etc.) and were not available for analysis. The
oblateness of the Earth causes the CGRO orbit to precess with a period
of about 53 days. During the precession cycle, high declination
sources ($|\delta|\ge41^\circ$)
%ones where the magnitude of declination exceeds 41
%degrees),
cease occulting near the orbital poles.

When occultation steps are measured, the background is estimated using
110 seconds of data before and after the occultation step, called the
fitting window. Occultation steps from sources other than the measured
source within this window are considered interfering sources. If
occultation steps of an interfering source occur within 10 seconds of
the measured source, that step is discarded. The number of
interfering sources in the fit affects the systematic error in three
ways (1) where sharing of flux occurs between source terms in the
fitting process and (2) where sources are either included or not
included in the fit because of an incorrect assumption about the
sources' relative intensity to the measured source or (3) when an
unknown source is present in the fitting window. Other sources of
systematic error include red noise and coherent pulses from bright
X-ray sources present, but not necessarily occulting, within the
fitting window.

To estimate the average systematic error, a grid covering the galactic
plane with points every 3 degrees along the galactic plane with two
additional sets of grid points at +6 degrees and -6 degrees galactic
latitude. Grid points within 2 degrees of bright occultation sources
where excluded, yielding a total sample of 156 points. The average
flux from each grid point was averaged over a 9 year period and the
standard deviation of 1-day flux averages was computed as a function
of galactic longitude. The values for the standard deviations are larger 
than
expected for a Gaussian distribution about zero flux due to various
systematic effects. We find broadening factors of about 30\%-60\% in
excess of normal statistics, strongest near the galactic center and
near bright variable sources such as Vela X-1 and Cygnus X-1. Both
positive and negative trends are present in the average fluxes. For
example, near EXO2030+375 ($l_{II}$ = 77.2 degrees, $b_{II}$ = -1.3
degrees), the average flux at $b_{II}$ = 0 degrees is 0.0077, 0.0009,
-0.001 photons cm$^{-2} s^{-1}$ and standard deviation 1.85, 1.42,
1.26 (1 sigma = 0.01 photons cm$^{-2} s^{-1}$) for longitudes $l_{II}$
= 72.0, 78.0, and 84.0 degrees, respectively (Harmon, B.A. et
al. 2002). Clearly interfering sources in various regions of the sky
produce both positive and negative offsets in flux. These fluxes may
be modulated at the precession period and may change sign, depending
on the limb geometry.

Data gaps arise due to three specific causes: passage through the South
Atlantic Anomaly when the detectors are switched off; known
interfering source occulting within 10 seconds of the target; periods
when the source is occulted by the earth's limb at too shallow an
angle for a significant detection to be made (or is not occulted at
all). The latter effect happens only for sources lying at greater than
41 degrees in declination.

	 Interfering sources that are positively identified
should be fitted for in the occultation technique, however this is
(as in the case of the background correction) a question of
optimisation and so interfering sources are not always completely
removed. A signature of source interference is significant negative
flux indicating that a persistent source close to the target has
not been fitted correctly in calculating the occultation step-height for
the target source. Provided the interfering source has not varied
on timescales comparable to the recurrence period of the target, it
will be averaged out in the epoch folding procedure. If however the
source has varied then its effects must be positively identified
before a folded lightcurve can be produced.

\begin{table*}
\caption{Summary of the observational parameters of HMXRBs studies in
this work.
A0538-668 is not included in the table because it was not detected in the 
BATSE
data sample. TJD = Julian date - 2440000.5. For EXO2030+375 epoch 1
corresponds to TJD8363.5 - 9353.5 and epoch 2 to  TJD10114.5 -
10723.4. Column 4 gives the epoch of periastron passage used (Bildsten et al 
1997), except in the case of 4U1145-619 and
4U1700-377 which are arbitary dates. The last two columns compare the
observed peak in the Lomb-Scargle power spectrum with Monte Carlo
simulated data - see text for details.}
\label{tab:summary}
\begin{tabular}{lcccccc}
&&&&& \multicolumn{1}{c}{BATSE data} &
  \multicolumn{1}{c}{Simulated data} \\
X-ray Source & Optical  & Orbital period  & T$_{0}$  & Detection
	      & Peak FWHM  & Peak FWHM  \\
              & Counterpart & (days) &  (TJD)  & Significance &
$\times 10^{-4} d^{-1}$ & $\times 10^{-4} d^{-1}$ \\

EXO2030+375 epoch 1         & Be & 46.01    & 8936.50 & 33.3 $\sigma$ & 8.70 
  & 9.06$\pm$0.34 \\
EXO2030+375 epoch 2         & Be & 46.01    & 8936.50 & 3.47  $\sigma$ & 
16.4  & 15.6$\pm$6.0 \\
4U1145-619  	      	     & Be & 186.5    & 8363.51 & 5.68 $\sigma$ & 4.56  & 
3.72$\pm$1.7 \\
4U1700-377                  & SG & 3.412    & 8363.54 & 241$\sigma$ & 3.50  
& 3.40$\pm$0.06 \\
4U1907+097                  & SG & 8.38     & 5578.75 & 3.41  $\sigma$ & 
4.86  & 3.61$\pm$2.0 \\
Cen X-3		     &    & 2.087    & 8561.50 & 29.7 $\sigma$ & 3.27  & 
3.44$\pm$0.13 \\
GX301-2	             & SG & 41.5     & 8802.79 & 139$\sigma$ & 3.65  & 
3.45$\pm$0.09 \\
LMC-X4		             & Be & 1.4, 30.35& 7741.99 & 86.0$\sigma$ & 3.34  & 
3.46$\pm$0.10 \\
OAO1657-41	             & SG & 10.44    & 8515.99 & 74.5 $\sigma$ & 3.26  & 
3.41$\pm0.09$ \\
\end{tabular}

%\medskip

%{\em 1} TJD8363.5 - 9353.5

%{\em 2} TJD10114.5 - 10723.4

%{\em 3} Epoch of periastron passage \cite{bil} used as zeropoint for
%folded lightcurves in Figure~\ref{fig:lightcurves}.

%{\em 4} Full-width-at-half-maximum of the peak in the Lomb-Scargle
%power spectrum corresponding to the X-ray modulation detected in the
%BATSE occultation data.

%{\em 5, 6} Values for simulated data are the mean and standard
%deviation calculated by Monte-Carlo simulation. }

\end{table*}

	Unidentified soures of interference pose an additional
problem in that their position relative to the earth's limb and
the target change continuously. This leads to modulation at the
precession period as the interference contribution varies cyclically
with the orbital plane.This type of artefact may be non sinusoidal
although subtracting a sinewave with correct period and amplitude
can attenuate it in some cases. In principle such signals could be
used iteratively to improve the occultation analysis but this is not
practical at this time. Persistent and identified transient sources
of interference can be investigated on a source by source basis.
In some cases datapoints or stretches of data have to be excluded,
if for example a nearby source  flares unpredictably. While in some
datasets the interference is clearly apparent as a completely
distinct feature in the power spectrum and can be attenuated by
subtracting a sinewave fit to the modulation if neccesary. Scargle
\shortcite{sca3} states that the modified power spectrum
(Lomb-Scargle periodogram) is equivalent to least-squares fitting with
sinewaves, hence our procedure is equivalent to filtering the data in
the frequency domain.

\subsection{Implications of uneven sampling}
Unevenly sampled data cannot be correctly analysed by the direct
Fourier Transform technique as the resulting power spectrum is a
convolution of both the source frequencies and sampling frequencies.
This leads to spurious frequencies and possible suppression of
genuine signals. Specifically; uneven sampling increases leakage of power
to nearby frequencies, although at the same time, aliasing may be reduced.
Our data are very unevenly sampled, especially
when we excluded subsets of the data e.g. data intervals
selected on the basis of an ephemeris in order to search for the
possible 16.65d transient modulation in A0538-668. Fortunately the
Lomb-Scargle periodogram is formulated for exactly this situation
and is relatively insensitive to the sampling structure (Scargle 1982, 
1989).

A further complication is the non-sinusoidal nature of many of the
lightcurves, generally the less sinusoidal a signal is, the more
components it will have in any frequency space representation. These
components are seen in power spectra as sidebands to the main
frequency spike. If the shape of an otherwise regular modulation
changes with time, these peaks can become broadened, shift and
become blended together. Thus in our period determinations we backed
up the Lomb-Scargle result with Phase Dispersion Minimisation
analysis (PDM). \cite{pdm}. Such periodograms typically contain a number
of local minima corresponding to (For example) integer multiples
of the true period. However the PDM technique is totally insensitive
to the shape of the lightcurve and this makes for highly accurate
period determinations in the case of irregular lightcurves. Periods were
measured from the PDM periodogram by calculating the centroid of the feature 
of
interest. We relied on the Lomb-Scargle periodogram for our primary 
detection tool
because the separation of Fourier components is better suited to
identification of interfering sources, sidebands, aliasing and
statistical modeling of the signal.  PDM was implemented in the Starlink 
program 'Period'.

\subsection[]{Evaluation of signal detection confidence levels}
In detecting weak, unevenly sampled, periodic signals against a varying
background, statistical significance is a key issue. The only reliable
way to evaluate the confidence levels associated with a power spectrum
under these conditions is by statistical simulation.

	We first measured the mean and variance of each dataset, we
then created a series of random numbers conforming to a Gaussian
distribution with the same mean and variance as each dataset. This
series was then combined with the sampling structure (Timing information)
of the original dataset by simply replacing each flux measurement with a
number drawn from the random gaussian distribution.
Thus creating a simulation of what BATSE might have
seen, had there been no periodic X-ray sources in its field of view.
This dataset was then subjected to Lomb-Scargle analysis and the power
of the highest peak and the highest peak in a specified frequency range 
recorded.
This process was then repeated 1000 times after which the mean and
and variance of the peak noise power (PNP)were calculated,
both for the whole spectral range available and for the specified frequency 
range
of interest. For the Lomb-Scargle periodograms presented here,
the displayed confidence levels are the mean PNP + 3 standard deviations.
Thus our confidence levels directly reflect how likely it is that a
given signal is real or due to chance without making any assumptions
about the underlying properties of the data. The confidence levels
calculated by this procedure for each dataset were later used in evaluating
the Lomb-Scargle periodograms generated from simulated datasets.

\subsection[]{Jitter in outburst turn-on times}

The width (FWHM) of peaks in the power spectrum may be related to
orbit-to-orbit variation in the exact time at which outbursts occur.
We investigated this by comparing the FWHM of the orbital modulation
peak in the Lomb-Scargle periodogram to that measured for a simulated
dataset containing a sinewave of appropriate period and amplitude
superimposed on a background created by randomizing the flux readings
in the original dataset.  Results are shown in Table~\ref{tab:summary}
for comparison.  Further investigation of this parameter by a
Monte-Carlo technique showed that the FWHM is dependent on the noise
present. Specifically the FWHM becomes more variable as the S/N ratio
deteriorates. Combined with the well known dependence of the FWHM on
the length of the dataset, the resolution of the periodogram in
respect of this parameter had to be calculated independently for each
dataset. For each source a Monte-Carlo simulation was run by first
finding the correct amplitude sinewave to give the same S/N ratio as
that seen in the data -using the limit finding procedure described in
section 3.1.1.  The flux values in the data were then repeatedly
randomized and added to this signal. The Lomb-Scargle periodogram was
calculated each time and the FWHM of the peak measured and
recorded. After 100 cycles the mean and standard deviation of the FWHM
were calculated and the results are displayed in
Table~\ref{tab:summary}.

In addition to movement in the outburst phase it is possible that
other effects such as variations in the shape of the outburst profile
might also give rise to a widening of the peak in the power
spectrum. However, inspection of the results presented in
Table~\ref{tab:summary} shows no evidence of any variability in the
power spectrum peak indicating a general lack of jitter in turn-on
times and no significant variation in the outburst profiles.

\section[]{Results}
\begin{figure*}
\begin{center}
\psfig{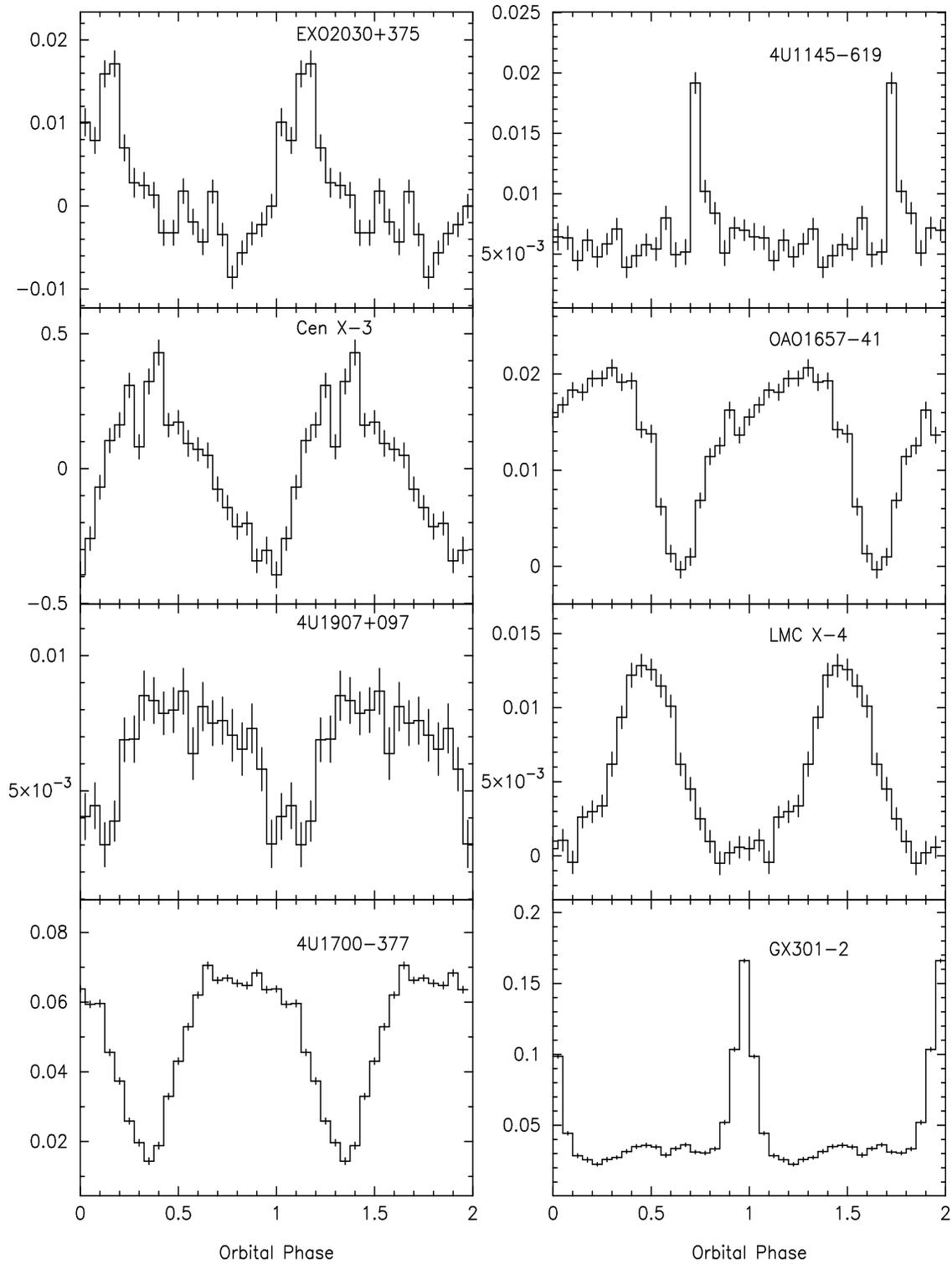}
\caption{Phase-averaged lightcurves created from BATSE occultation
flux measurements. All data are folded at the periods and phases given
in Table~\ref{tab:summary} except for 4U1907+09 which is folded at
twice the period quoted in the table (see text for details).}
\label{fig:lightcurves}
\end{center}
\end{figure*}

Table~\ref{tab:summary} summarises key observational parameters of the
HMXBs studied, while Figure~\ref{fig:lightcurves} shows the X-ray
lightcurves folded over 20 phase-bins. The errors shown are
observational errors inherent in the occultation analysis, in
producing the averaged lightcurves, the error-bars are the standard
deviation of points in each phase-bin.

\begin{figure*}
\begin{center}
\psfig{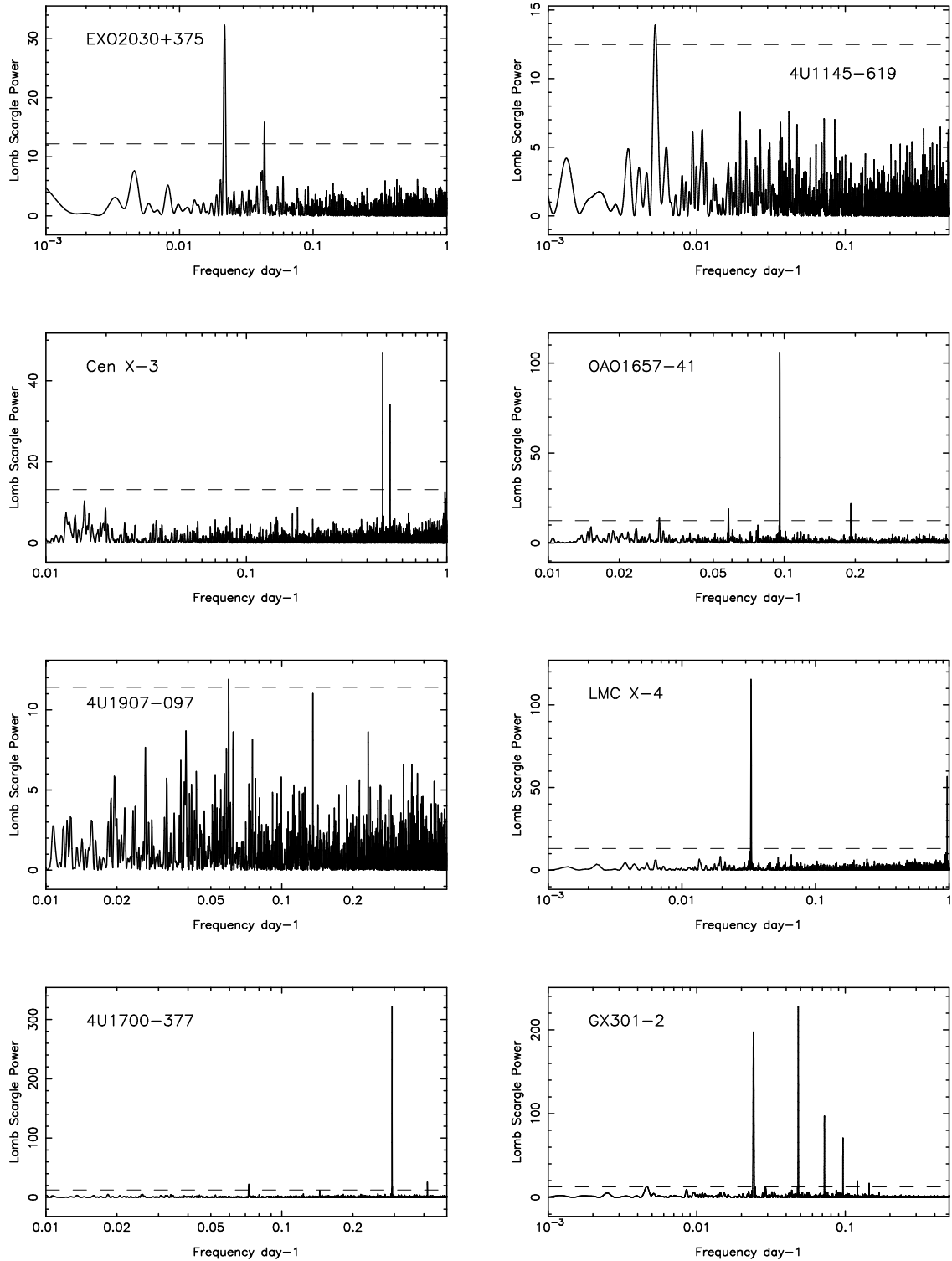}
\caption{Lomb-Scargle periodograms of BATSE occultation flux histories.
The 3$\sigma$ significance levels are plotted as dashed lines.
These levels are calculated by Monte-Carlo simulations over the
frequency space displayed (see Section 2.2).}
\label{fig:lsdata}
\end{center}
\end{figure*}

\subsection{ Link between Type I X-ray outbursts and long period
optical variations in A0538-668 }
We tested the prediction that A0538-668 only exhibits X-ray outbursts
correlated with its orbital period at times coincident with minima in
its long period optical lightcurve. \cite{al}
We note that previous widely spaced X-ray observations of A0538-668 tend
to support this hypothesis, however we find 
conclusively that between
TJD 8363 - 10999 (The time span of the BATSE occultation data) no 
significant
modulation of 20-70 keV X-ray flux at the orbital period is present at the 
predicted epochs.
To do this we calculated an ephemeris for precise times at which this model
predicts outbursts to occur. The orbital parameters were taken from
Skinner \shortcite{sk}, P=16.65d, with a zeropoint T$_{0}$=TJD 3423.96 
Maxima are expected at
phase 0.9 in this scheme if the X-rays follow
the optical variability. The long period parameters from Alcock et al 
\shortcite{al}
are under this scheme: P=420.82d, with arbitrary zeropoint TJD 9001. The 
long period
minima occur at arbitrary phase 0.55.
The ephemeris was then used to isolate time intervals when X-ray outbursts 
were expected. The
duration of the time interval examined either side of long period minima was 
100 days
to cover the data coincident with the optical lightcurve "dip" regions.

According to Skinner \shortcite{sk}, A0538-668 was in the field of view of 
the Einstein satellite
on four occasions. According to the ephemeris described here,  detections 
reported on
16 Dec 1980 and 3 Feb 1981 lie in the optical dip, and non detections on 10 
Apr 1979 and 30 May
1980 lie firmly in the centre of the bright phase of the optical lightcurve.

\subsubsection[]{BATSE detection threshold for A0538-668}
By creating test datasets with the same mean and variance as the
BATSE data and containing only simulated signals, we determined how
strong a periodic signal must be in order to give a 3$\sigma$ detection.
This approach could also be used to model other components of a
power spectrum, particularly where interfering sources and sampling
structure lead to spurious frequencies.

Known interfering signals in the A0538-668 data include LMC X-4
(which is only 0.6$\degr$ away and dominates the periodic flux in this 
dataset)
and the spacecraft precessional modulation. These components were modelled
by measuring their amplitudes in folded lightcurves at 35.5d and 52d
respectively.

Using the ephemeris described above BATSE data were
selected which were coincident with the four long-period optical
minima observed during the MACHO observations of Alcock et al 
\shortcite{al}.
This resulted in four blocks of BATSE one-day averages each of duration 
$~$100 days
spanning the time interval TJD 9124.52 - TJD 10999.5
Having failed to detect any statistically significant
modulation consistent with 16.65 days, this subset of the BATSE observations
of A0538-668 was then repeatedly randomized to destroy any periodic
content and combined with an incrementaly increasing sinusoidal
signal plus fixed sinewave components to simulate the interfering sources
in the real data. After each
step, the Lomb-Scargle periodogram was computed and
the highest peak in a pre-determined frequency range was compared to
the 3 $\sigma$ confidence level previously calculated for that
frequency range as described in section 2.3.  Once this threshold was
exceeded, the amplitude of the sine wave signal was recorded.  The
detectability of such a weak signal is somewhat influenced
by the noise present. At the very margins of detectability the power
in a given frequency can fluctuate up and down despite a gradual
increase in signal amplitude.  Hence we used the criterion that the
power in the selected frequency range must remain above the threshold
value irrespective of chance structures in noise in order to
qualify. The derived upper limit was independently tested using
many different noise sets to verify this criterion was met, i.e. That
a signal of the amplitude quoted would definitely have been detected
by the methods described here, had it been present.  Weaker signals
would not consistently rise above the threshold although
detections of marginal significance are possible.

\section[]{Discussion}
\subsection[]{EXO2030+375}

This object has been subject of several recent studies. Reig et al 
\shortcite{rei}
presented results of multi-wavelength observations and modeling
of the accretion rate. Wilson et al (2002) report
on a decade of X-ray observations combined with IR and H$\alpha$
measurements which show a decline in the density of the circumstellar
disk around the Be star. This decline was followed by a sudden drop in
the X-ray flux and a turn-over from a spin-up trend to spin-down in
the frequency history. This is the first Be/X-ray binary to show an
extended interval, about 2.5 years, where the global trend is
spin-down, but the outbursts continue. In 1995 the orbital phase of
EXO 2030+375's outbursts shifted from peaking about 6 days after
periastron to peaking before periastron.  The outburst phase slowly
recovered to peaking at about 2.5 days after periastron. Wilson et al
(2002) interpret this shift in orbital phase followed by a slow
recovery as evidence for a global one-armed oscillation propagating in
the Be disk. This was further supported by changes in the shape of the
H$\alpha$ profile which are commonly believed to be produced by a
reconfiguration of the Be disk. The truncated viscous decretion disk
model provided an explanation for the long series of normal outbursts
and the evidence for an accretion disk in the brighter normal
outbursts.

In the present analysis we find convincing evidence for elevated flux levels
at apastron in the earlier epoch (see Figure~\ref{fig:lightcurves}).
The RXTE ASM lightcurve for epoch 2 was studied by Reig et al 
\shortcite{rei},
we also analysed this data in order to compare the lightcurve and power 
spectrum
with the BATSE results. The power spectra from both instruments appear to 
share
the same features in epochs 1 \& 2, peaks at $P_{orb}$ \& $P_{orb}/2$ with 
powers
at similar ratios. Direct comparison of the BATSE and RXTE folded 
lightcurves suggests
that X-ray emission from EXO2030+375 may be softer during the apastron
outbursts than during the normal Type I outbursts at periastron. This
hypothesis is drawn from the relative amplitudes of the two peaks in
the lightcurve and the differing spectral responses of the two
instruments.

Comparison of the FWHM of the orbital period peak at 46d with that
derived from simulated data reveals that although the phase of peak
outburst intensity shifts between the two epochs, its recurrence is
otherwise periodic.  There is no evidence for any gradual shift in
phase in either epoch.

\subsection[]{A0538-668}

We find that during the period coincident with the MACHO monitoring of
Alcock et al \shortcite{al}, A0538-668 did not exhibit hard X-ray
outbursts at the 16.65 day orbital period during either the optical
minima or maxima. The upper limit for such activity is found to be 3.0
$\times 10^{-3}$ photon cm$^{-2}$ s$^{-1}$at 20-70 kev. We compare
this limit with the X-ray spectrum obtained with Ariel V \cite{wc} at
the peak of the July 1977 outburst. By assuming a mean photon energy
of 45 keV over a spectral range of 20-70 keV for the BATSE observations
we estimate a maximum flux density of 2.7 $\times 10^{-3}$ keV
cm$^{-2}$ s$^{-1}$ keV$^{-1}$. The fitted spectrum of White \&
Carpenter \shortcite{wc} is extremely steep and dominated by a thermal
component. By extrapolating their model, our lower limit for detection
by BATSE in the data presented here is consistent with the flux that
would be expected on the basis of this model to within errors. Thus we
conclude that had A0538-668 been active at its previous level BATSE
would have detected it.

\subsection[]{Centaurus X-3}

This source was detected in BATSE data at 29$\sigma$ with a period
2.087d. Cen X-3 is the only HMXB to have a well determined orbital
period derivative \cite{nag} however this is too small to show up in
our analysis.  The power spectrum in Figure~\ref{fig:lsdata} shows two
peaks, but comparison with the power spectrum for simulated data showed
the higher-frequency peak to be an artefact of aliasing. The folded
lightcurve Figure~\ref{fig:lightcurves} is a simple sawtooth like
form, brightening more rapidly than it decays.

\subsection[]{4U1907+097}

Two peaks with $\sim$3$\sigma$ significance
are found at periods of 16.8d and 7.39d.  The
orbital period of the system is believed to be 8.38d (Marshall and
Ricketts 1980, Cook and Page 1987) thus the 16.8d X-ray modulation is
consistent with twice the binary period.  However, there is no evidence for 
a
peak in the power spectrum at $\sim$8.4d - this is not surprising
if one considers the folded lightcurve in Figure 1 which shows just a
single broad peak when folded at 16.8d. Constructing a simulated
dataset containing a sinewave at 8.38d with amplitude taken as 0.002
(measured from a lightcurve folded at this period) failed to reproduce
the 16.8d modulation. Thus it appears that the emission is really
modulated at twice the binary period in these BATSE data.

\subsection[]{4U1700-377}

Early BATSE occulation data from this source have already been the
subject of a paper by Rubin et al., 1996.  Our PDM period agrees with
3.41180$\pm$0.00003d from Copernicus X-ray observation \cite{bran}. We
also detect significant peaks in the power spectrum at 13.808d at a
significance of 11.3$\sigma$, and at exactly half this period. This
period has been previously reported by Konig \& Maisack
\shortcite{km97}.  We further investigated this period by subtracting
the orbital modulation from the lightcurve, using
Figure~\ref{fig:lightcurves} as a template. This procedure resulted in
reduction of spectral power at the orbital period by a factor of
24. No attenuation of the 13.808d period was observed, in fact a
significant increase resulted. Hence we are confident that the 13.808d
period is an independent modulation and not an harmonic of the orbital
period.

\subsection[]{4U1145-619}

A peak was detected in LS periodogram at 190.238d with FWHM=17.48
days. A PDM periodogram determination of 186.6d $\pm$0.348068d
measured by centroiding is consistent with the established value of
186.5 days.
%Peaks of marginal significance are detected corresponding
%to half the binary period and interference at the CGRO precession
%period.

\subsection[]{OAO1657-41}

The PDM period of 10.4626$\pm$0.0011d is in agreement
with the period of 10.4436$\pm$0.0038d \cite{ch1} from BATSE pulse
timing studies.

\subsection[]{LMC X-4}

The 1.4d binary period is not detected. The "average" Nyquist period of
the daily average data is 2.4d so the single-step data
was also analysed, again with no detection. The 30.5d
X-ray modulation believed to be due to the precession of a warped
accretion disk \cite{lang} is detected and is shown to be persistent
throughout the full span of the data Figure~\ref{fig:lightcurves}. All
period finding techniques tried, (Lomb-Scargle, PDM, CLEAN) derive a
period for this modulation slightly shorter than usually quoted. A PDM
periodogram constructed at a resolution of 2$\times 10^{-5}$ cycles
and measured by centroiding, derived a period of 30.349d while the L-S
periodogram yielded 30.355d. Monte Carlo simulations of sinusoidal
signals at 30.5d and 30.35d superimposed on noise created by
randomizing the original dataset show that the LS periodogram
frequency resolution is 1.43765$\times 10^{-5}$ at the S/N level seen
in the data. A solution at 30.5d is thus seemingly in conflict with
the BATSE dataset. A remarkable fact is the stability of this
modulation over the 7 years of data, since no jitter is evident from
the values in Table~\ref{tab:summary}

\subsection[]{GX301-2}

This system is clearly the brightest, most regular source studied
here. The phase-binned folded lightcurve shows increasing flux levels
consistent with phase 0.5, each of which reached type-I outburst
levels. This result confirms previous work by Pravdo et al, 1995 and
Koh et al, 1997 - both of these works also present evidence for
apastron outbursts from this system.
The power spectrum for GX301-2 shows a pattern of strong peaks
which after comparison to simulated data are found to be real.  These
frequency components, appearing at integer submultiples of the orbital
period are a signature of the highly non-sinusoidal lightcurve. The
spectrum shown is that generated after the removal of 4 particularly
bright apastron flux points, as expected their inclusion boosts the
$P_{orb}/2$ component.

\section{Conclusions}

Observations of accreting pulsars in high mass binary systems using
the BATSE occultation technique reveal that this approach is
sufficiently sensitive to derive accurate measurements of the X-ray
modulation. Evidence for apastron X-ray outbursts is found in 2
sources: EXO2030+375 and 4U1145-619. All of these sources are
in high mass, long period systems, EXO2030+375 and 4U1145-619 have
similarities in their folded lightcurves having flux levels
significantly above minimum for most of each orbit and apastron
features reaching ~30-40 percent of peak flux. GX301-2 has a much
flatter minimum level between emissions and the apastron emission is
only ~10 percent of peak flux. These differences undoubtedly reflect
the different nature of the companion star, whether Be or
supergiant. The fact that none of the shorter period systems show
evidence for this phenomenon may be explainable in terms of the
standard Bondi-Hoyle accretion model if suitable parameters can be
estimated/calculated for the outflow velocity and density of the
circumstellar material/wind along the orbital trajectory of the
neutron star. We hypothesise that in short period systems the radial
profile of the stellar wind can never satisfy the required conditions
for apastron outbursts due to the smaller apastron separation
distances and higher neutron star velocities involved.  The FWHM of
the orbital period peaks in the power spectra were investigated as a
possible indicator of variability in the exact time at which outbursts
"switch on". To within the uncertainties imposed by the signal/noise
ratio of the data this effect was not detected in any of the sources.

\section*{Acknowledgments}

SL acknowledges the use of STARLINK software and the support of a PPARC 
studentship.

\label{lastpage}

\end{document}